\documentclass[12pt, final,onecolumn, romanappendices]{IEEEtran}
\usepackage{amsmath,amssymb}
\usepackage[bookmarks,colorlinks]{hyperref} 
\usepackage{acronym}  
\usepackage{color}  
\usepackage{soul}
\usepackage{graphicx,psfrag,subfigure,cite}
\usepackage{mathtools}

\usepackage{bbm}
\usepackage{bigints}
\usepackage{mathrsfs,cancel}
\usepackage[table]{xcolor}
\usepackage{algorithm}
\usepackage[end]{algpseudocode}
\usepackage{arydshln}

\usepackage{tikz}
\usepackage{pgfplotstable}
\usepackage{rotate}

\definecolor{rubblue}{cmyk}{1,0.5,0,0.6}
\definecolor{rubgreen}{cmyk}{0.5,0,1,0}
\definecolor{rubgray}{cmyk}{0.03,0.03,0.03,0.1}

\usepgfplotslibrary{units}

\usetikzlibrary{%
patterns,%
calc,%
fit,%
arrows,%
plotmarks,%
shadows,%
chains,%
shapes%
}

\tikzset{>=latex'} 
\tikzstyle{every picture}+=[remember picture] 
\pgfdeclarelayer{background}
\pgfdeclarelayer{foreground}
\pgfsetlayers{background,main,foreground}

\tikzstyle{blueblock}=[draw=rubblue, rectangle, thick, drop shadow, minimum width=20mm, minimum height=8mm,fill=rubblue!20, text width=20mm, text centered]
\tikzstyle{bluebox}=[draw=rubblue, rectangle, thick, drop shadow, minimum width=8mm, minimum height=8mm,fill=rubblue!20, text width=8mm, text centered]
\tikzstyle{greenblock}=[draw=rubgreen, rectangle, thick, drop shadow, minimum width=20mm, minimum height=8mm,fill=rubgreen!20, text width=20mm, text centered]
\tikzstyle{dot} = [draw, circle, minimum size=0.2pt,scale=0.3,fill=black,black]
\tikzstyle{smalldot} = [draw, circle, minimum size=0.1pt,scale=0.2,fill=black,black]
\tikzstyle{reddot}  =[draw,circle,minimum size=0.2pt,scale=0.8,fill=red,thin]
\tikzstyle{greendot}  =[draw,circle,minimum size=0.2pt,scale=0.8,fill=Green,thin]
\tikzstyle{bluedot}  =[draw,circle,minimum size=0.2pt,scale=0.8,fill=blue,thin]
\tikzstyle{whitedot}=[draw,circle,minimum size=0.2pt,scale=0.8,fill=white,thin]
\tikzstyle{blackdot} = [draw, circle, minimum size=0.2pt,scale=0.7,fill=black,black]
\tikzstyle{sum} = [drop shadow, draw=rubblue, thick, fill=rubblue!20, circle]
\tikzstyle{relay} = [blueblock, minimum width=5mm, minimum height=20mm, text width=5mm, rounded corners=2pt]
\tikzstyle{relay2} = [blueblock, minimum width=5mm, minimum height=15mm, text width=5mm, rounded corners=2pt]
\tikzstyle{relay3} = [blueblock, minimum width=5mm, minimum height=25mm, text width=5mm, rounded corners=2pt]
\tikzstyle{relay4} = [blueblock, minimum width=5mm, minimum height=10mm, text width=5mm, rounded corners=2pt]
\tikzstyle{relay5} = [blueblock, minimum width=5mm, minimum height=50mm, text width=5mm, rounded corners=2pt]
\tikzstyle{relay6} = [blueblock, minimum width=5mm, minimum height=5mm, text width=5mm, rounded corners=2pt]
\tikzstyle{circgreen} = [draw, circle, inner sep=2pt, fill=rubgreen, drop shadow, thick]
\tikzstyle{circwhite} = [draw, circle, inner sep=2pt, fill=white, drop shadow, thick]
\tikzstyle{circdashed} = [draw, dashed, circle, inner sep=2pt, fill=rubgray, drop shadow, thick]
\tikzstyle{vertbox} = [rectangle, draw=rubblue, thick, rotate=90, text centered, minimum width=16.5mm, minimum height=8mm, text width=16.5mm, inner sep=0pt, fill=rubblue!20, drop shadow]
\tikzstyle{vertboxb} = [rectangle, draw=rubblue, thick, rotate=90, text centered, minimum width=16.5mm, minimum height=8mm, text width=16.5mm, fill=rubblue!20, drop shadow]
\tikzstyle{vertboxshort} = [rectangle, draw=rubblue, thick, rotate=90, text centered, minimum width=10mm, minimum height=8mm, text width=10mm, inner sep=0pt, fill=rubblue!20, drop shadow]
\tikzstyle{smalldotgreen} = [draw=rubgreen, circle, minimum size=0.2pt,scale=0.8,fill=rubgreen!20]
\tikzstyle{antenna} = [regular polygon, regular polygon sides=3, draw, shape border rotate=180, minimum size=0.2pt, scale=0.3]

\tikzstyle{poly} = [regular polygon, regular polygon sides=6, shape aspect=0.5, minimum width=1.5cm, minimum height=0.35cm, draw, dashed]

\definecolor{cff9e00}{RGB}{255,158,0}
\definecolor{c4fff00}{RGB}{79,255,0}
\definecolor{cff0012}{RGB}{255,0,18}
\definecolor{c00c5ff}{RGB}{0,197,255}
\definecolor{c046f00}{RGB}{4,111,0}
\definecolor{c004b9d}{RGB}{0,75,157}

\newlength{\mylen}
\usetikzlibrary{petri}
\usetikzlibrary{shapes}
\usetikzlibrary{positioning,arrows,patterns}
\usepackage[utf8x]{inputenc}
\usepackage{scalefnt}
\usetikzlibrary{calc,decorations.markings}
\pgfplotsset{compat=1.10}
\usetikzlibrary{arrows.meta}

\usepackage[columnwise,switch,mathlines]{lineno} 

\usepackage{pgfplots}
\pgfplotsset{compat=newest}
\usepgfplotslibrary{groupplots}

\usepackage{bm}

\DeclareMathAlphabet{\mathsfbr}{OT1}{cmss}{m}{n}
\SetMathAlphabet{\mathsfbr}{bold}{OT1}{cmss}{bx}{n}
\DeclareRobustCommand{\msf}[1]{%
  \ifcat\noexpand#1\relax\msfgreek{#1}\else\mathsfbr{#1}\fi
}

\makeatletter
\newcommand{\msfgreek}[1]{\csname s\expandafter\@gobble\string#1\endcsname}
\makeatother

\DeclareFontEncoding{LGR}{}{} 
\DeclareSymbolFont{sfgreek}{LGR}{cmss}{m}{n}
\SetSymbolFont{sfgreek}{bold}{LGR}{cmss}{bx}{n}
\DeclareMathSymbol{\salpha}{\mathord}{sfgreek}{`a}
\DeclareMathSymbol{\sbeta}{\mathord}{sfgreek}{`b}
\DeclareMathSymbol{\sgamma}{\mathord}{sfgreek}{`g}
\DeclareMathSymbol{\sdelta}{\mathord}{sfgreek}{`d}
\DeclareMathSymbol{\sepsilon}{\mathord}{sfgreek}{`e}
\DeclareMathSymbol{\szeta}{\mathord}{sfgreek}{`z}
\DeclareMathSymbol{\seta}{\mathord}{sfgreek}{`h}
\DeclareMathSymbol{\stheta}{\mathord}{sfgreek}{`j}
\DeclareMathSymbol{\siota}{\mathord}{sfgreek}{`i}
\DeclareMathSymbol{\skappa}{\mathord}{sfgreek}{`k}
\DeclareMathSymbol{\slambda}{\mathord}{sfgreek}{`l}
\DeclareMathSymbol{\smu}{\mathord}{sfgreek}{`m}
\DeclareMathSymbol{\snu}{\mathord}{sfgreek}{`n}
\DeclareMathSymbol{\sxi}{\mathord}{sfgreek}{`x}
\DeclareMathSymbol{\somicron}{\mathord}{sfgreek}{`o}
\DeclareMathSymbol{\spi}{\mathord}{sfgreek}{`p}
\DeclareMathSymbol{\srho}{\mathord}{sfgreek}{`r}
\DeclareMathSymbol{\ssigma}{\mathord}{sfgreek}{`s}
\DeclareMathSymbol{\stau}{\mathord}{sfgreek}{`t}
\DeclareMathSymbol{\supsilon}{\mathord}{sfgreek}{`u}
\DeclareMathSymbol{\sphi}{\mathord}{sfgreek}{`f}
\DeclareMathSymbol{\schi}{\mathord}{sfgreek}{`q}
\DeclareMathSymbol{\spsi}{\mathord}{sfgreek}{`y}
\DeclareMathSymbol{\somega}{\mathord}{sfgreek}{`w}

\DeclareMathSymbol{\svarsigma}{\mathord}{sfgreek}{`c}

\DeclareMathSymbol{\sGamma}{\mathalpha}{sfgreek}{`G}
\DeclareMathSymbol{\sDelta}{\mathalpha}{sfgreek}{`D}
\DeclareMathSymbol{\sTheta}{\mathalpha}{sfgreek}{`J}
\DeclareMathSymbol{\sLambda}{\mathalpha}{sfgreek}{`L}
\DeclareMathSymbol{\sXi}{\mathalpha}{sfgreek}{`X}
\DeclareMathSymbol{\sPi}{\mathalpha}{sfgreek}{`P}
\DeclareMathSymbol{\sSigma}{\mathalpha}{sfgreek}{`S}
\DeclareMathSymbol{\sUpsilon}{\mathalpha}{sfgreek}{`U}
\DeclareMathSymbol{\sPhi}{\mathalpha}{sfgreek}{`F}
\DeclareMathSymbol{\sPsi}{\mathalpha}{sfgreek}{`Y}
\DeclareMathSymbol{\sOmega}{\mathalpha}{sfgreek}{`W}

\DeclareRobustCommand{\mcal}[1]{%
  \ifcat\noexpand#1\relax\mathnormal{#1}\else\cal{#1}\fi
}
\DeclareRobustCommand{\BM}[1]{%
  \ifcat\noexpand#1\relax\bm{\boldUppercaseItalicGreek{#1}}\else\bm{#1}\fi
}
\makeatletter
\newcommand{\boldUppercaseItalicGreek}[1]{\csname var\expandafter\@gobble\string#1\endcsname}
\makeatother

\newcommand{\V}[1]{\bm{#1}}
\newcommand{\M}[1]{\BM{#1}}

\newcommand{\beq}{\begin{equation}}
\newcommand{\eeq}{\end{equation}}

\newtheorem{proposition}{Proposition}
\newtheorem{corollary}{Corollary}
\newtheorem{lemma}{Lemma}

\newtheorem{remark}{Remark}

\definecolor{myblack}{rgb}{0,0,0}
\definecolor{myblack2}{rgb}{0,0,0}
\newcommand{\paperTitle}{Characterizing IoT Networks with Asynchronous Time-Sensitive Periodic Traffic}
\newcommand{\paperTitleMarkboth}{Characterizing IoT Networks with Asynchronous  Time-Sensitive Periodic Traffic}

\acrodef{IoT}{Internet of things}
\acrodef{TSP}{transmission success probability}
\acrodef{SIR}{signal-to-interference ratio}
\acrodef{MC}{{Markov chain}}

\definecolor{mygreentext}{RGB}{31,97,25} 
\definecolor{myredtext}{RGB}{207,21,24}
\definecolor{mybluetext}{RGB}{0,43,187} 
\definecolor{myblacktext}{RGB}{0,0,0} 
\definecolor{mygray}{RGB}{200,200,200} 
\definecolor{myorange}{RGB}{255, 178, 102}
\definecolor{mymagenta}{rgb}{0.78, 0.08, 0.52}
\definecolor{mymyblacktext}{rgb}{0, 0.74, 1}

\newcommand{\savespaces}{\vspace{-4pt}}

	\algnewcommand{\Initialize}[1]{%
	\State \textbf{Initialize:}
	\Statex \hspace*{\algorithmicindent}\parbox[t]{.8\linewidth}{\raggedright #1}
}
\begin{document}




\title{\paperTitle}


\author{
	\vspace{0.2cm}
       Hesham~ElSawy,~\IEEEmembership{Senior~Member,~IEEE} 
        \thanks{H.\ ElSawy is with the Electrical Engineering Department, King Fahd University of Petroleum and Minerals (KFUPM), Dhahran, Saudi Arabia, (e-mail: \texttt{hesham.elsawy@kfupm.edu.sa}).
        The author acknowledges the support received from the deanship of scientific research (DSR) at KFUPM under grant no. DF191052.}
}
\maketitle 
\markboth{Arxiv Version}{ElSawy: \paperTitleMarkboth}


\setcounter{page}{1}

\begin{abstract}
This paper develops a novel spatiotemporal model for large-scale IoT networks with asynchronous periodic traffic and hard-packet deadlines. A static marked Poisson bipolar point process is utilized to model the spatial locations of the IoT devices, where the marks mimic the relative time-offsets of traffic duty cycles at different devices. {At each device, an absorbing Markov chain is utilized to capture the temporal evolution of packets from generation until either successful delivery or deadline expiry.} The temporal evolution of packets is defined in terms of the Aloha transmission/backoff states. From the network perspective, the meta distribution of the transmission success probability is used to characterize the mutual interference among of the coexisting devices. To this end, the network performance is characterized in terms of the probabilities of meeting/missing the delivery deadlines and transmission latency. The results unveil counter-intuitive superior performance of strict packet deadlines in terms of transmission success and latency.
\end{abstract}


\acresetall		

\savespaces
\section{Introduction}\label{sec:intro}

   The \ac{IoT} promotes ubiquitous connectivity that bridges the physical and cyber worlds. The \ac{IoT} is foreseen to enable better monitoring and smarter automation to several verticals that include industrial, smart-grids, agricultural, transportation, healthcare, and public safety domains. Indeed, each sector has its unique traffic patterns as well as reliability and latency requirements~\cite{URLLC}. Out of the several classes that may exist, this paper focuses on large-scale \ac{IoT} networks with time-sensitive periodic traffic, which may appear in monitoring systems that require timely transmissions (i.e., before a deadline) of periodic updates~\cite{URLLC, Hard_deadline}. 
   
   Spatiotemporal models have been recently developed to jointly account for both the spatial network topology and the temporal traffic flow~\cite{First_Elsawy}. From the temporal perspective, queueing theory is utilized to account for the packets arrivals/departures as well as for devices' status and activities. From the spatial perspective, stochastic geometry is utilized to account for the mutual interactions (e.g., interference, spectrum access, and contention) among the active devices. For instance, the scalability and stability of uplink networks and ad-hoc networks with unsaturated traffic are characterized in~\cite{GhaElsBadAlo:17, Chisci, Martin_queue}. Latency for grant-free and grant-based uplink access are assessed and contrasted  in~\cite{Gharbieh}. Latency for different downlink scheduling schemes are compared in \cite{Tony}. Random access for massive IoT networks is studied in~\cite{Deng}. Prioritized traffic in large-scale ad-hoc network is characterized in~\cite{Nardelli}. Age-of-information for different traffic patterns is characterized in~\cite{AoI}. Latency of uplink non-orthogonal multiple access is studied in~\cite{Sheng}. Self-sustainability of energy harvesting IoT networks is studied in \cite{Fatma}. However, all of the aforementioned efforts ignore the effect of packets deadlines.  
   
   To the best of the author's knowledge, the spatiotemporal characterization of large-scale ad hoc networks with periodic traffic and packet-deadlines is still an open problem. Motivated by~\cite{URLLC, Hard_deadline}, this paper focuses on such an important category.


\section{System Model}\label{sec:intro}

\subsubsection{\textbf{Network Model}} A time-slotted system is considered, where each transmitter generates a packet every $T$ time slots. However, the duty cycles for packets generation at all transmitters in the network are not necessarily synchronized. {To account for space and time, the IoT network is modeled via a static marked Poisson bipolar point process (PBP) $(\mathrm{\Psi},\Delta) \subset \mathbb{R}^2 \times \{0,1,2,\cdots, T-1\}$. That is, transmitters are spatially distributed according to an arbitrary realization of a PPP $\mathrm{\Psi} \subset \mathbb{R}^2$ with intensity $\lambda$. Each transmitter $\V{\chi}_i \in \mathrm{\Psi}$ has a single receiver that is randomly located on the circumference of a circle centered at $\V{\chi}_i$ with radius $R$.} To model the asynchronous traffic generation, each transmitter $\V{\chi}_i$ is assigned an independent discrete time mark $\Delta_{\V{\chi}_i}$ that is uniformly distributed in the range $[0,T-1]$. The relative time offsets for the duty cycles of two transmitters $\V{\chi}_i$ and $\V{\chi}_j$ is given by $|\Delta_{\V{\chi}_i}-\Delta_{\V{\chi}_j}|$, where $|\cdot|$ denotes the absolute value.

\subsubsection{\textbf{Time sensitivity}} Each packet is tagged with a life-time $\tau \in \{\tau_{\rm min},\tau_{\rm min}+1,\cdots,T-1\}$, determined in number of time slots, to represent hard transmission deadlines. Hence, $\tau=\tau_{\rm min}$ represents the most strict deadline in the traffic. The packets lifetime is decremented by one each time slot and a packet is discarded when its lifetime elapses. Deadlines are assumed to be less than the traffic period $T$.\footnote{{The constraint $\tau<T$ implies no accumulation of packets at the devices, which simplifies the analysis. Such assumption applies to monitoring applications where the receivers are interested in the most recent measurements, system status, or updates.}} For the sake of generality, deadlines are assumed to vary across  packets such that $\tau$ is drawn from a discrete distribution $f_{\tau}(\tau)$ in the range of $\tau_{\rm min}\leq \tau \leq T-1$. The deadlines of all packets are independent and identically distributed. The special case of equal deadlines $\tau_o$ for all packets is naturally captured as $f_{\tau}(\tau)\!=\!\mathbbm{1}_{\{\tau=\tau_o\}}$, where $\mathbbm{1}_{\{\cdot\}}$ is the indicator function.

\subsubsection{\textbf{Communication model}} {Packet arrivals occur at the beginning of the time slot}. Transmitters with non-empty buffers, denoted hereafter as active transmitters, follow an Aloha protocol with parameter $p_A$. At each time slot, an active transmitter selects to transmit with probability $p_A$ or to defer transmission to subsequent time slot with probability $\bar{p}_A=(1-p_A)$, hereafter the bar notation $\bar{(\cdot)}=1-(\cdot)$ is used for probability complement. The Aloha parameter $p_A$ balances the tradeoff between aggregate network interference and transmission latency. Packets are transmitted at a constant power level of $w$ mW. The power of transmitted signals decays at the rate $r^{-\eta}$ with the distance $r$, where $\eta>2$ is the path-loss exponent. The signals are also subject to unit mean exponentially distributed power fading, which is independent across different time slots and different locations.

Active transmitters persistently follow the Aloha protocol until either successful transmission or elapsed deadline. In the former case, the \ac{SIR} at the receiver should exceed the threshold $\theta$ for successful decoding. In the latter case, the message becomes obsolete, and hence, is discarded from the device's buffer. Consequently, a transmitter returns to the idle state due to either successful packet delivery ($\text{SIR}_t > \theta$) or elapsed deadline ($t>\tau$). 

\subsubsection{\textbf{Time Indexing}}

 {Without loss of generality, for each device in the network, local and relative time indexing is utilized to indicate a certain time slot position within a given duty cycle $nT$, $\forall$ $n\in\mathbbm{Z}$. For any device, the local time slot $t=1$ denotes the starting point of its own duty cycle where packets are generated and the $1^{\rm st}$ channel access is attempted. Similarly, the local time slot $t= 3$ denotes two time slots later where the $3^{\rm rd}$ channel access attempt may occur. Note that, in the local time slots $t \in \{2,\cdots,T-1\}$ the transmitter is active if and only if the generated packet has experienced $t$ consecutive unsuccessful attempts (i.e., due to backoff or SIR violation) and has deadline $\tau>t$. It is worth noting that devices with similar (different) time offsets will have aligned (misaligned) local time slot incidences.}  

\section{Analysis}

For the sake of organized presentation, a microscopic single-device model is first presented. Such single-device model is then extended to the macroscopic network wide model.  

\subsubsection{\textbf{Single-Device Model}}

\begin{figure}[t]
	~~~~~~~~~~~~~~~~~~~~~~~~~~~{
		\begin{tikzpicture}[fill=blue,ultra thick, scale=0.8, transform shape,font=\large]
		\input{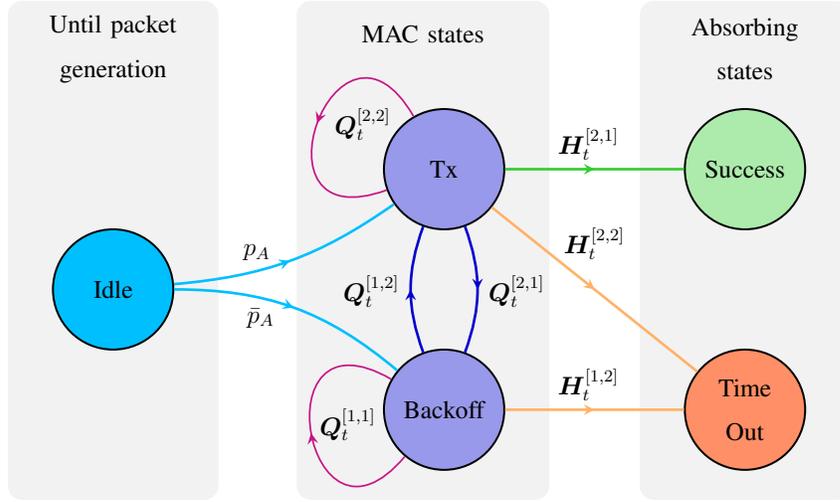}
		\end{tikzpicture}}
	\caption{The discrete time absorbing  \ac{MC} for temporal packet evolution from generation to absorption. The superscript in rectangular parentheses $[i,j]$ is used to determine the element in the $i$-th row and $j$-th column of a matrix. }
	\label{fig:MC2}
\end{figure} 
The states evolution of the Aloha protocol at each transmitter and final packet status can be described via an absorbing \ac{MC} as shown in Fig.~\ref{fig:MC2}. {As shown in \cite{Hae:J16, Chisci}, the \ac{TSP} of packets vary across devices according to the aggregate interference experienced at their locations. Consequently, we categorize the coexisting links according to the experienced \ac{TSP} to $L$ equal percentiles, denoted hereafter as \ac{TSP} classes.}  Consider a device that belongs to the $\ell^{\rm th}$ \ac{TSP} class, then the absorbing \ac{MC} in Fig.~\ref{fig:MC2} is mathematically represented by the following transition matrix~\cite{alfa}
\begin{linenomath}
	\beq \label{eq:AMC0}
	\M{P}^{(\ell)} =\left[\begin{matrix}
		\;\; \M{Q}^{(\ell)} & \M{H}^{(\ell)} \;\;\\ 
	\;\;	\M{0} \;\;\;\; &  \M{I} \;\;\;\; \;\;          	
	\end{matrix}\right],
	\eeq
\end{linenomath}
where $	\M{Q}^{(\ell)}$ is a sub-stochastic matrix describing transitions among transient states, $\M{H}^{(\ell)}$ is a sub-stochastic matrix describing the transition to absorbing states due to either successful transmission or elapsed deadlines, and $\M{I}$ is the identity matrix of size $2\times 2$ representing the two absorbing states. Hereafter, the superscript $(\ell)$ is used to denote the TSP class of a device. The matrix $\M{Q}^{(\ell)}$ is of size $2T \times 2T$ and the matrix $\M{H}^{(\ell)}$  is of size $2T\times2$ as defined by the duty cycle, the two-states Aloha protocol, and the two absorbing states. To capture the MAC protocol states and to tract the elapsed time in transmission, $	\M{P}^{(\ell)} $ in \eqref{eq:AMC0} can be structured as follows  
\begin{equation}\label{eq:AMC2}
\!\!\! \M{P}^{(\ell)} \!\!=\!\! \left[
\begin{array}{cccccc;{2pt/2pt}c}
\M{0}   & 	\M{Q}^{(\ell)}_1  & \M{0}  & \M{0}  &\cdots & \M{0}  &  \M{H}^{(\ell)}_{1}         \\
\M{0}   & \M{0} & \M{Q}^{(\ell)}_2  & \M{0}& \cdots & \M{0} & \M{H}^{(\ell)}_{2}          \\  
\vdots &  \vdots & \ddots & \ddots & \ddots & \vdots & \vdots  \\
\M{0}   & \M{0} & \M{0}  & \M{0}& \cdots & \M{Q}^{(\ell)}_{T-2} & \M{H}^{(\ell)}_{T-2}  \\ 
\M{0}   & \M{0} & \M{0}  & \M{0}& \cdots & \M{0} & \M{H}^{(\ell)}_{T-1}          \\  
\hdashline[2pt/2pt]
\M{0}   & \M{0} & \M{0}  & \M{0}& \cdots & \M{0} & \M{I} 
\end{array}
\right],
\end{equation}
where the dash lines in \eqref{eq:AMC2} partition $\M{P}^{(\ell)}$ to the form in \eqref{eq:AMC0}. The rows in \eqref{eq:AMC2} depict the progressing time evolution of the packet until absorption. That is, all the matrices $\M{Q}^{(\ell)}_t$ are in the upper off-diagonal positions to capture the progressing time evolution. The matrix $\M{Q}^{(\ell)}_t$ depicts the protocol evolution between time-slots $t$ and $t+1$. {Hence, the two-states of the Aloha protocol are further divided into $T-1$ logical Aloha states to capture the local time slot index.}  Each of the matrices $\M{H}^{(\ell)}_t$ is of size $2 \times 2$ to capture the probability of absorption, from each MAC state, due to successful transmission or elapsed deadline. The detailed structure of $\M{Q}^{(\ell)}_t$ and $\M{H}^{(\ell)}_t$ are given by 

\begin{linenomath}
	\beq \label{eq:Ti}
\!\!	\M{Q}^{(\ell)}_t \!\!=\!\! \bar{F}_\tau(t) \left[\begin{matrix}
	\bar{p}_A  &  p_A \\
	 \bar{s}_\ell \bar{p}_A  &   \bar{s}_\ell p_A
\end{matrix}\right] \quad \!\!\!\! \text{and} \!\! \quad 	\M{H}^{(\ell)}_t \!\!=\!\! \left[\begin{matrix}
		0   & 	F_\tau(t)  \\
		s_\ell  &  \bar{s}_\ell F_\tau(t) 
	\end{matrix}\right],
	\eeq
\end{linenomath}
where $s_\ell=\mathbb{P}({\rm SIR}>\theta)$ is the probability to meet the decoding threshold for a device belonging to the $\ell^{\rm th}$-\ac{TSP} class and $F_\tau(t) = \sum_{t=0}^{t} f_{\tau}(t)$ is the cumulative density function (CDF) of the deadline distribution. As shown from \eqref{eq:Ti}, $\M{Q}^{(\ell)}_t$, $\forall t$,  captures the transient Aloha evolution when a transmission failure occurs (i.e., with probability $\bar{s}_\ell$) and the deadline is not elapsed (i.e., with probability $\bar{F}_\tau(t)=1-F_\tau(t)$). In contrast, $\M{H}^{(\ell)}_t$, $\forall t$, captures the two absorption events where transmission success occurs (i.e., with probability ${s}_\ell$) or the deadline elapses (i.e., with probability \textbf{ ${F}_\tau(t)$}).

The transient solution of the absorbing \ac{MC} shown in Fig.~\ref{fig:MC2}, and described by \eqref{eq:AMC2} and \eqref{eq:Ti}, is characterized via the following lemma
\begin{lemma} \label{Lem_1}
{Let the vector $\V{\mathrm{x}}^{(\ell)}_t=\{x^{(\ell)}_{0,t}, x^{(\ell)}_{1,t}\}$ define the probabilities that a device belonging to the $\ell^{\rm th}$ TSP class is at the backoff state (index $0$) or transmission state (index $1$) at the $t^{\rm th}$ local time slot. Also, let  $\V{\mathrm{y}}^{(\ell)}_t=\{y^{(\ell)}_{s,t}, y^{(\ell)}_{f,t}\}$, where $y^{(\ell)}_{s,t}$ and $y^{(\ell)}_{f,t}$ are the probabilities of absorption to, respectively, success and timeout states at time slot $t$ for an $\ell^{\rm th}$-TSP class device.} Then,  

\begin{equation}
\!\!\!\!\!\!\!\!\!\!\!\!\V{\mathrm{x}}^{(\ell)}_t =\left\{\begin{matrix}
\V{\beta}; &  \quad t=1 \\
&   \\
\V{\beta} \times \prod_{i=1}^{t-1} \M{Q}^{(\ell)}_i; &  \quad 1< t \leq T-1 \\
&   \\
\V{0} &    t=T
\end{matrix}\right.    	
\end{equation} 
and
{\begin{equation}
\V{\mathrm{y}}^{(\ell)}_t = 
\left\{\begin{matrix}
\V{0}  & \quad t=1 \\
&   \\
\V{\beta} \times \M{H}^{(\ell)}_{1}; & \quad t=2 \\
&   \\
\V{\beta} \times \left(\prod_{i=1}^{t-2} \M{Q}^{(\ell)}_i\right) \times \M{H}^{(\ell)}_{t-1}; &   \quad 2 < t \leq T
\end{matrix}\right.   
\end{equation}}
where $\V{\beta}=\{\bar{p}_A, p_A\}$ is the initialization vector of the Aloha protocol and both  $\M{Q}^{(\ell)}_i$ and $\M{H}^{(\ell)}_i$ are defined in \eqref{eq:Ti}.
\end{lemma}
\begin{IEEEproof}
	According to \eqref{eq:AMC0}, at each time slot, the probability of staying within the transient states is captured by $\M{Q}^{(\ell)}$ and the probability of absorption is captured by  $\M{H}^{(\ell)}$. Hence, the probability of staying in  transient for $t$-consecutive time slots is given by $\left(\M{Q}^{(\ell)}\right)^t$. Similarly, the probability of being absorbed after exactly $t$ time slots is $\left(\M{Q}^{(\ell)}\right)^t \times \M{H}^{(\ell)}$. Exploiting the structures of  $\M{Q}^{(\ell)}$ and  $\M{H}^{(\ell)}$ shown in \eqref{eq:AMC2} and identifying the Aloha protocol initialization with $\V{\beta}$, the lemma is proved. 
	\end{IEEEproof}

\begin{remark} \label{Rem_1}
{Both $\V{\mathrm{x}}^{(\ell)}_t$ and $\V{\mathrm{y}}^{(\ell)}_t$ are sub-stochastic vectors that satisfy the following condition  $\V{\mathrm{x}}^{(\ell)}_t\times \mathbf{1} + \sum_{i=1}^t \V{\mathrm{y}}^{(\ell)}_i\times \mathbf{1} = 1$, where $\V{1}$ is a $2\times 1$ column vector of ones. This is because, at each time slot, the device is either i) attempting to transmit its packet by following the Aloha protocol or ii) idle due to absorption in any of the preceding time slots $i< t$ (i.e., with probability $\sum_{i=1}^t \V{\mathrm{y}}^{(\ell)}_i\times \mathbf{1}$).} Furthermore, the latency distribution of successfully transmitted packets can be calculated as $\mathbb{P}(\text{latency}=t) ={y^{(\ell)}_{s,t-1}}\times ({\sum_ty^{(\ell)}_{s,t}})^{-1}$. 
\end{remark}

The performance of the absorbing \ac{MC} in \eqref{eq:AMC2} is characterized in the following corollary
\begin{corollary}
Let $\V{\mathrm{a}}^{(\ell)}=\{a^{(\ell)}_s,a^{(\ell)}_f\}$ denote the final state of a generic packet for a device in the $\ell^{\rm th}$ TSP class, where $a^{(\ell)}_s$ and $a^{(\ell)}_f=1-a^{(\ell)}_s$ are, respectively, the probability of being eventually absorbed due to transmission success and elapsed deadline. Also, Let $\V{\mathrm{d}}^{(\ell)}=\{d^{(\ell)}_s, d^{(\ell)}_f\}$ denote a scaled mean time to adsorption for an $\ell$-TSP class device, where $\frac{d^{(\ell)}_s}{a^{(\ell)}_s}$ and $\frac{d^{(\ell)}_f}{a^{(\ell)}_f}$ are, respectively, the mean time to absorption due to transmission success and elapsed deadline. Then,
\begin{equation}
\V{\mathrm{a}}^{(\ell)}= \beta \times \left( \M{H}^{(\ell)}_1 + \sum_{i=2}^{T-2} \left(\prod_{t=1}^{i-1} \M{Q}^{(\ell)}_{t}\right) \times \M{H}^{(\ell)}_i  \right),  
\end{equation}  
and
\begin{equation}
\V{\mathrm{d}}^{(\ell)}= \beta \times \left( \M{H}^{(\ell)}_1 +  \sum_{i=2}^{T-2} i\; \left(\prod_{t=1}^{i-1} \M{Q}^{(\ell)}_{t}\right) \times \M{H}^{(\ell)}_i    \right). 
\end{equation}

\end{corollary}
\begin{IEEEproof}
Applying the law of total probability, accounting for the $(T-1)$-time slot maximum life-time of the packet, and following the same arguments as in Lemma~\ref{Lem_1}, the corollary is proved.
\end{IEEEproof}

\subsubsection{\textbf{Network-wide Model}} 
The determinism of the periodic traffic generation along with the static locations and fixed time offsets of devices impose a location-dependent \ac{TSP} that is maintained across different duty cycles. {Consider an arbitrary network-wide (i.e., global) modulo $T$ indexing for the system time slots. Due to the different local time offsets across the devices, a randomly selected global time slot $t_g\in\{1, 2,  \cdots, T-1\}$ will accommodate transmission attempts of the following devices; i)  1$^{\rm st}$ transmission attempts of all devices with time offset $\Delta=t_g$, ii) 2$^{\rm nd}$ transmission attempts of devices with one leading time offset $\Delta=t_g-1$ that experienced one unsuccessful transmission attempt; iii) 3$^{\rm rd}$ transmission attempt of devices with two leading time offset $\Delta=t_g-2$ that experienced two consecutive  unsuccessful transmission attempts; and so on.} Since the time offset distribution is uniform and location independent, then all global time slots will accommodate similar transmission statistics. Averaging over all devices classes and time offsets of the duty cycles, we obtain the following time-slot and TSP class independent activity and absorption probability vectors 
\begin{equation} \label{averaged}
 \V{\mathrm{x}} \!=\!\! \sum_{\Delta=1}^{T-1} \frac{\mathbb{E}_{\ell} \left\{\V{\mathrm{x}}^{(\ell)}_\Delta\right\}}{T-1} \quad \text{and} \quad  \V{\mathrm{y}} \!=\!\! \sum_{\Delta=1}^{T-1} \frac{\mathbb{E}_{\ell} \left\{\V{\mathrm{y}}^{(\ell)}_\Delta\right\}}{T-1},
\end{equation}
where $\V{\mathrm{x}}=\{x_0,x_1\}$, $\V{\mathrm{y}}=\{y_s,y_f\}$, and $\V{\mathrm{x}} \times \V{1} + \V{\mathrm{y}} \times \V{1} =1$. Regardless of the time slot index,  $\V{\mathrm{x}}$ in \eqref{averaged} provides the probability that a randomly selected device is active and utilizing one of the two states of the Aloha protocol. Similarly, $\V{\mathrm{y}}$ in \eqref{averaged} provides the probability that a randomly selected device has been absorbed in any of the preceding time-slots due to either successful transmission or deadline time-out. 

By virtue of \eqref{averaged} and exploiting the mean filed effect of the aggregate interference, we can conduct an approximate time-slot independent \ac{TSP} analysis. Such approximation is then validated in Section~IV. Consider a randomly selected bipolar link with transmitter located at $\V{\chi}_\circ$ and receiver located at $\V{\nu}_\circ$. Using \eqref{averaged}, we define the point process of potential interfering devices $\tilde{\mathrm{\Psi}}$ with intensity $(1-y_s) \lambda$. The point process $\tilde{\mathrm{\Psi}}$ excludes, from the aggregate interference, the devices that are always successful (i.e., idle) before the test bipolar link is active. Then, \ac{TSP} of such bipolar link is  
\begin{align}
\!s_{\V{\chi}_\circ} \!=\! \mathbb{P}\left\{\frac{w h_\circ R^{-\eta}}{\sum_{\V{\chi}_j  \in \tilde{\mathrm{\Psi}} \setminus \V{\chi}_\circ} \mathbbm{1}_{\{e_{j}\}} w\; g_j \; \|\V{\chi}_j- \V{\nu}_\circ\|^{-\eta}}\! >\! \theta\; \big| \mathrm{\Psi} \right\} \notag \\
\overset{(*)}{=} \prod_{\tilde{\mathrm{\Psi}}\setminus \V{\chi}_\circ} \left( {   \frac{x_{1} }{(1+\theta \frac{  R^{\eta} }{\|\V{\chi}_j - \V{\nu}_\circ\|^{\eta} }) (1- y_s)}+ \frac{{x_{0}} +  y_f}{1-y_s}}\right),
\label{sinr1}
\end{align}
where $h_\circ$  is the useful channel gain, $e_j$ is the event that the $j^{\rm th}$ potential interfering device is active (i.e., not in backoff or timeout), $g_j$ is the $j^{\rm th}$ interfering channel gain. Note that $(*)$ is obtained by averaging over the channel gains and devices activities, where the probabilities $x_1$, $x_0$ and $y_f$ are normalized with respect to $(1-y_s)$ to provide a legitimate distribution for the three possible states of the potential interfering devices.

To compute the expectation in \eqref{averaged}, the distribution of $s_\ell$ across all links in the network is required. Such distribution, denoted as the meta distribution of the \ac{TSP} is defined as:
\begin{equation} \label{meta1}
\bar{F}_{s}(\theta,\gamma ) = \mathbb{P}\left\{ \mathbb{P}\left\{ \rm{SIR} > \theta \big| \mathrm{\Psi} \right\} > \gamma \right\} = \mathbb{P}\left\{ s > \gamma \right\}.
\end{equation} 
For a given SIR threshold $\theta$, the meta distribution in \eqref{meta1} captures the percentile of links that achieve \ac{TSP} greater than $\gamma$. 
For analytical tractability~\cite{Hae:J16}, the meta distribution of \ac{TSP} is usually approximated as shown in the following proposition. 
\begin{proposition} \label{lem2}
	The meta distribution of the \ac{TSP} is approximated as 
	\begin{align} \label{meta2}
	&\bar{F}_{s}(\theta,\gamma)\! \approx \!  1\!-\! \mathcal{I}_\gamma \left(\frac{M_1 (M_1-M_2)}{M_2- M_1^2},\frac{(1-M_1) (M_1-M_2)}{M_2-M_1^2} \right)\!\!,
	\end{align}
	where $\mathcal{I}_\gamma(a,b) = \frac{1}{\mathcal{B}(a,b)}\int_0^\gamma t^{a-1} (1-t)^{b-1} {\rm d}t$ is the regularized incomplete beta function, $M_1$ and $M_2$ are the first two moments of the \ac{TSP} given by 
	\begin{align} \label{mom1}
	M_1 
	& =  \text{exp}\left\{ -\frac{2\pi^2 \lambda R^2 x_1 \theta^{\frac{2}{\eta}}}{\eta \sin(2\pi/\eta)}  \right\},
	\end{align}
	and
		\begin{align}  \label{mom2}
	\!\!\!\!M_2 
	\!\!= \!\text{exp}\left\{\!-\frac{2\pi^2 \lambda x_{1} \theta^\frac{2}{\eta} R^2}{\eta^2 \sin (2\pi/\eta)}    \left(\!2\eta  \!-\! \frac{(\!\eta\!-\!2) x_{1}}{1-y_s}\right)   \right\}.
	\end{align}
	\begin{IEEEproof}
		The approximation is obtained as in \cite{Hae:J16}. Starting from \eqref{sinr1}, an integral form for the moments $M_1$ and $M_2$ can be obtained using the probability generating functional of the PPP. Applying change of variables and exploiting the state probabilities in~\eqref{averaged}, the integral can be solved and the proposition is proved. 
	\end{IEEEproof}
\end{proposition}
 {For mathematical convenience, we discretize the meta distribution in \eqref{meta2} to $L$ equal percentiles of TSP classes. Let $\omega_0=0$ and $\omega_L=1$, then define the set  $\{\omega_2,\omega_3, \cdots, \omega_{L-1}\}$ such that 
 \begin{align} \label{meta3}
 \bar{F}_s(\theta,\omega_\ell)-\bar{F}_s(\theta,\omega_{\ell-1}) =\frac{1}{L}.
 \end{align}
The \ac{TSP}s of all devices within the range   $[\omega_{\ell},\omega_{\ell+1}]$ are approximated via the median value $s_\ell$, which is given by
\begin{align} \label{meta4}
\bar{F}_{s}(\theta,\omega_\ell)\!-\!\bar{F}_{s}(\theta,s_\ell) \!=\!\bar{F}_{s}(\theta,s_\ell)\!-\!\bar{F}_{s}(\theta,\omega_{\ell-1}) \!=\! \frac{1}{2 L}.
\end{align}
Note that increasing $L$ shrinks the interval $[\omega_{\ell},\omega_{\ell+1}]$. Hence, improving the accuracy of $s_\ell$ in representing the TSP within each class.}

\subsubsection{\textbf{Overall spatiotemporal model}} Proposition 1 and Lemma 1 are both interwoven. The meta distribution of the TSP in Proposition 1 requires the state probabilities derived through Lemma 1. Meanwhile, Lemma 1 require the TSP for each class obtained through the discretization step after Proposition 1. By virtue of the fixed point theory, such interdependency can be resolved as summarized in Algorithm~\ref{alg:succprobmarkdisc}.

\begin{algorithm}[!t]
	\caption{Spatiotemporal Model}
	\label{alg:succprobmarkdisc}

	\begin{algorithmic}[1]
		\Require $\lambda$, $\eta$, $\theta$, and $\V{\beta}$.
			\State \parbox[t]{\dimexpr\linewidth- \algorithmicindent * 4}{\textbf{Initialize:} $i\leftarrow1$, $\V{\mathrm{y}}^{[1]} \leftarrow \V{0}$,  $\V{\mathrm{x}}^{[1]} \leftarrow \frac{\V{\beta}}{T-1}$
			\strut}
		\While {$\text{max}(\V{\mathrm{x}}^{[i]} - \V{\mathrm{x}}^{[i-1]})>\epsilon$}
	\Statex	\textbf{Stochastic geometry Analysis:}  
		\State \parbox[t]{\dimexpr\linewidth- \algorithmicindent * 4}{Use $x_1 \in \V{\mathrm{x}}^{[i]}$ and $y_s \in \V{\mathrm{y}}^{[i]}$, to construct  $\bar{F}_{s}(\theta,\gamma)$ as in Proposition \ref{lem2};\strut}  
		\For{each TSP class $\ell=1,2,\dots,L-1$}
		\State \parbox[t]{\dimexpr\linewidth- \algorithmicindent * 4}{Retrieve the values of ${s}_\ell$ using \eqref{meta3} \& \eqref{meta4}. \strut}  
	\EndFor
		\Statex	\textbf{Queueing Analysis:}
		\For {each time slot $t=1,2,\dots,T-1$} 
				\For{for each TSP class $\ell=1,2,\dots,L-1$}
		\State \parbox[t]{\dimexpr\linewidth- \algorithmicindent * 4}{Use $s_\ell$ \& \eqref{eq:Ti} to construct  $\M{Q}^{(\ell)}_t$ \& $\M{H}^{(\ell)}_t$;\strut}
			\State \parbox[t]{\dimexpr\linewidth- \algorithmicindent * 4}{Obtain   $\V{\mathrm{x}}^{(\ell)}_{t+1}$ and $\V{\mathrm{y}}^{(\ell)}_{t+1}$  as:
			\beq \nonumber
			\V{\mathrm{x}}^{(\ell)}_{t+1}=	\V{\mathrm{x}}^{(\ell)}_{t} \times \M{Q}^{(\ell)}_t \quad \text{and} \quad \V{\mathrm{y}}^{(\ell)}_{t+1}=	\V{\mathrm{x}}^{(\ell)}_{t} \times \M{H}^{(\ell)}_t  
			\eeq   
			\strut}
		\EndFor
			\State \parbox[t]{\dimexpr\linewidth- \algorithmicindent * 4}{Average over the TSP classes as:
			\beq \nonumber
			\mathbb{E}\left\{\V{\mathrm{x}}^{(\ell)}_{t+1}\right\}=	\sum_{\ell=1}^{L} \frac{\V{\mathrm{x}}^{(\ell)}_{t+1}}{L}  \quad \text{and} \quad 	\mathbb{E}\left\{\V{\mathrm{y}}^{(\ell)}_{t+1}\right\}=	\sum \frac{\V{\mathrm{y}}^{(\ell)}_{t+1}}{L}   
			\eeq   
			\strut} 
		\EndFor 
			\State \parbox[t]{\dimexpr\linewidth- \algorithmicindent * 4}{Obtain    $\V{\mathrm{x}}$  and $\V{\mathrm{y}}$ as in \eqref{averaged};\strut} 
		\State \parbox[t]{\dimexpr\linewidth- \algorithmicindent * 4}{ $i\leftarrow i+1$ \& $\mathbf{x}^{[i]}\leftarrow \{x_0, x_1\}$
			\strut}
		\EndWhile\\
		\textbf{Return} $\M{Q}^{(\ell)}_t$ and $\M{H}^{(\ell)}_t$, $\forall \ell, t$ for further computations.
	\end{algorithmic}
\end{algorithm}

\section{Numerical Results}\label{sec:conclude}
This section first verifies the developed spatiotemporal  model against independent Monte Carlo simulations. A single realization of a marked Poisson bipolar network with intensity $\lambda=0.05$ device/m$^2$ and time offset mark $\Delta\in\{0,T-1\}$ is realized in $(350 \times 350)$ m$^2$ area with wrap-around boundaries. A uniform distribution for deadlines is selected, a detection threshold $\theta=5$ is employed, and {$L=25$ TSP classes} are utilized. Each simulation run represents a time-slot where 1) packets are generated at designated devices, 2) channels gains are realized, 3) SIRs at active devices are computed, 4) packets departures are recorded, and 5) deadlines are traced. Devices that successfully transmit their packets or have elapsed timeout are kept idle until the start of their next duty cycle.

\begin{figure}[!t]
	\centering
	\includegraphics[trim={1cm 0.5cm 1.2cm 0.2cm},clip, width=0.6\linewidth]{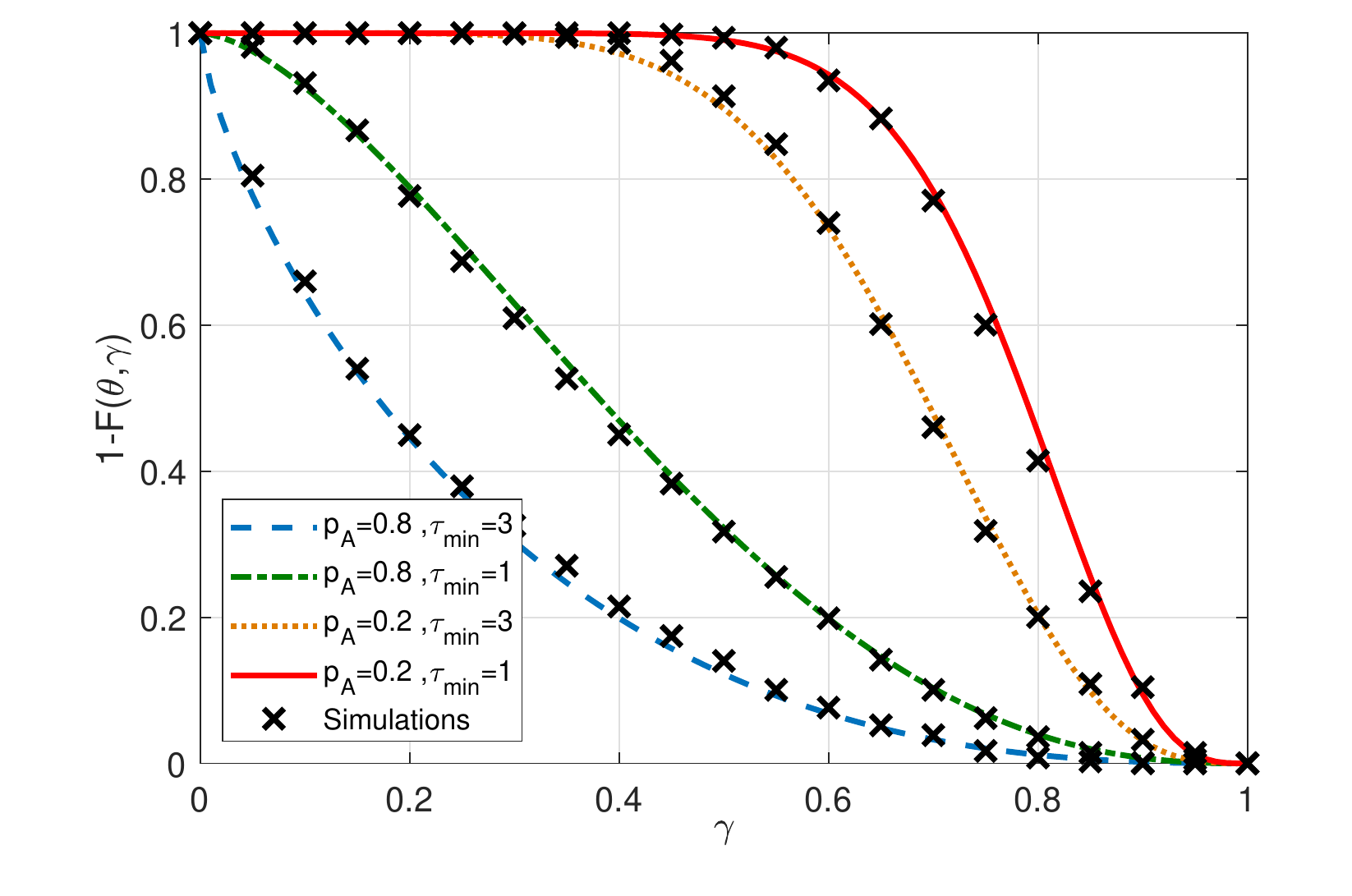}
	\caption{Meta distribution of the TSP $T=4$, $\lambda=0.05$, and $R=2$.}
	\label{fig:valid}
\end{figure}


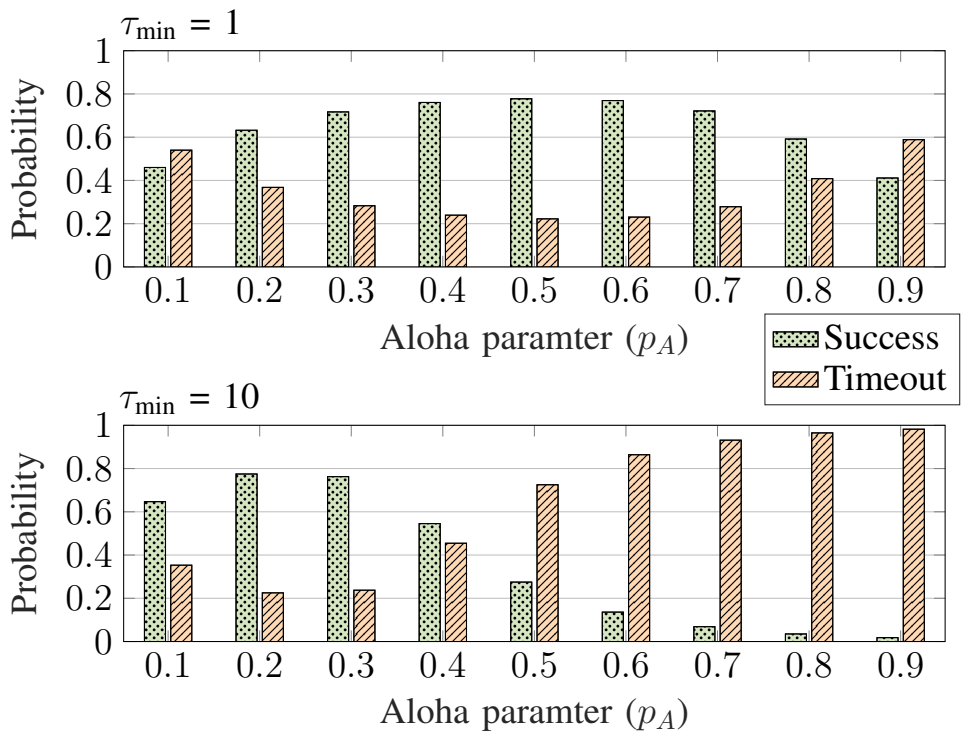
\begin{figure}[t]
\centering
	{\begin{tikzpicture}[scale=0.99, transform shape,font=\large]
%
%

\definecolor{mycolor1}{rgb}{0.46667,0.67451,0.18824}%

\begin{axis}[%
width=4.343in,
height=1.144in,
at={(0.729in,2.282in)},
scale only axis,
bar shift auto,
xmin=0.0514285714285714,
xmax=0.948571428571429,
ymajorgrids,
xlabel style={font=\color{white!15!black}},
xlabel={\large{Aloha paramter ($p_A$)}},
ymin=0,
ymax=1,
ylabel style={font=\color{white!15!black}},
ylabel={\large{Probability}},
axis background/.style={fill=white},
title style={at={(0.07,0.93)}, font=\large},
title={\large{$\tau_{\text{min}}\text{ = 1}$}},
legend style={at={(0.78,-0.64)}, anchor=south west, legend columns=1, legend cell align=left, align=left, draw=white!15!black}
]
\addplot[ybar, bar width=0.023, fill=mycolor1!30!white, postaction={pattern=crosshatch dots},  area legend, draw=black] table[row sep=crcr] {%
0.1	0.459938800349617\\
0.2	0.632074558775469\\
0.3	0.716954980960009\\
0.4	0.76059880757599\\
0.5	0.777645121653762\\
0.6	0.769457841687017\\
0.7	0.721570455400679\\
0.8	0.591669112613834\\
0.9	0.411345758765905\\
};

\addlegendentry{Success}

\addplot[ybar, bar width=0.023, fill=orange!30!white, postaction={pattern=north east lines}, area legend, draw=black] table[row sep=crcr] {%
0.1	0.540061199650384\\
0.2	0.367925441224532\\
0.3	0.283045019039991\\
0.4	0.23940119242401\\
0.5	0.222354878346238\\
0.6	0.230542158312983\\
0.7	0.278429544599321\\
0.8	0.408330887386166\\
0.9	0.588654241234095\\
};
\addlegendentry{Timeout}

\end{axis}

\begin{axis}[%
width=4.343in,
height=1.144in,
at={(0.729in,0.3in)},
scale only axis,
bar shift auto,
xmin=0.0514285714285714,
xmax=0.948571428571429,
ymajorgrids,
xlabel style={font=\color{white!15!black}},
xlabel={\large{Aloha paramter ($p_A$)}},
ymin=0,
ymax=1,
ylabel style={font=\color{white!15!black}},
ylabel={\large{Probability}},
axis background/.style={fill=white},
title style={at={(0.08,0.93)}, font=\large},
title={\large{$\tau_{\text{min}}\text{ = 10}$}},
legend style={at={(-4.464,-4.443)}, anchor=south west, legend columns=1, legend cell align=left, align=left, draw=white!15!black}
]
\addplot[ybar, bar width=0.023,fill=mycolor1!30!white, postaction={pattern=crosshatch dots}, area legend, draw=black] table[row sep=crcr] {%
0.1	0.646916804838457\\
0.2	0.774862101834046\\
0.3	0.762624865118082\\
0.4	0.545410633266355\\
0.5	0.274929461156037\\
0.6	0.136095475581622\\
0.7	0.0687081975261585\\
0.8	0.0350840550013155\\
0.9	0.0179756182908999\\
};

\addplot[ybar, bar width=0.023, fill=orange!30!white, postaction={pattern=north east lines}, area legend, draw=black] table[row sep=crcr] {%
0.1	0.353083195161543\\
0.2	0.225137898165955\\
0.3	0.237375134881918\\
0.4	0.454589366733645\\
0.5	0.725070538843963\\
0.6	0.863904524418378\\
0.7	0.931291802473841\\
0.8	0.964915944998684\\
0.9	0.9820243817091\\
};

\end{axis}
		\end{tikzpicture}}
	\caption{Absorption probabilities vs $p_A$ at $T=50$, $\lambda=0.5$, and $R=2$.}
	\label{fig:absorb}
\end{figure}


\begin{figure}[t]
	\centering
	\!\!\!\!\!\!\!\!\!{\begin{tikzpicture}[scale=0.99, transform shape,font=\large]
%
%
\definecolor{mycolor1}{rgb}{0.46667,0.67451,0.18824}%
\definecolor{mycolor2}{rgb}{0.49020,0.18039,0.56078}%

\begin{axis}[%
width=4.343in,
height=1.144in,
at={(0.729in,2.282in)},
scale only axis,
bar shift auto,
xmin=0.0514285714285714,
xmax=0.948571428571429,
xlabel style={font=\color{white!15!black}},
xlabel={\large{Aloha paramter ($p_A$)}},
ymin=0,
ymax=8,
ymajorgrids,
ylabel style={font=\color{white!15!black}},
ylabel={\large{Mean delay}},
axis background/.style={fill=white},
title style={at={(0.07,0.93)}, font=\large},
title={\large{$\tau{}_{\text{min}}\text{ = 1}$}},
legend style={at={(0.78,-0.63)}, anchor=south west, legend columns=1, legend cell align=left, align=left, draw=white!15!black}
]
\addplot[ybar, bar width=0.023, fill=mycolor1!30!white, postaction={pattern=crosshatch dots},  area legend, draw=black] table[row sep=crcr] {%
0.1	4.73779792671314\\
0.2	4.14398768526299\\
0.3	3.7618509004265\\
0.4	3.51852692473979\\
0.5	3.39289121629058\\
0.6	3.39768839239065\\
0.7	3.59813566262938\\
0.8	4.10946047585017\\
0.9	4.68779818776211\\
};
\addlegendentry{Success}

\addplot[ybar, bar width=0.023,fill=orange!30!white, postaction={pattern=north east lines}, area legend, draw=black] table[row sep=crcr] {%
0.1	7.15304985631904\\
0.2	6.37783256082743\\
0.3	5.88484855658174\\
0.4	5.60587115040636\\
0.5	5.52096009955185\\
0.6	5.63907116843963\\
0.7	6.01435706299078\\
0.8	6.71843900506539\\
0.9	7.40285567075322\\
};
\addlegendentry{Timeout}

\end{axis}

\begin{axis}[%
width=4.343in,
height=1.144in,
at={(0.729in,0.3in)},
scale only axis,
bar shift auto,
xmin=0.0514285714285714,
xmax=0.948571428571429,
xlabel style={font=\color{white!15!black}},
xlabel={\large{Aloha paramter ($p_A$)}},
ymin=0,
ymax=20,
ymajorgrids,
ylabel style={font=\color{white!15!black}},
ylabel={\large{Mean delay}},
axis background/.style={fill=white},
title style={at={(0.08,0.93)}, font=\large},
title={\large{$\tau{}_{\text{min}}\text{ = 10}$}},
legend style={at={(0.465,0.442)}, anchor=south west, legend columns=2, legend cell align=left, align=left, draw=white!15!black}
]
\addplot[ybar, bar width=0.023, fill=mycolor1!30!white, postaction={pattern=crosshatch dots}, area legend, draw=black] table[row sep=crcr] {%
0.1	7.42425449145411\\
0.2	6.65671561393506\\
0.3	6.61500780343942\\
0.4	7.60677789366361\\
0.5	8.47271765676039\\
0.6	8.83732289274535\\
0.7	9.01116252571482\\
0.8	9.10381657194331\\
0.9	9.15666130857833\\
};

\addplot[ybar, bar width=0.023, fill=orange!30!white,  postaction={pattern=north east lines}, area legend, draw=black] table[row sep=crcr] {%
0.1	15.7412411670282\\
0.2	15.4228754660754\\
0.3	15.5085356027673\\
0.4	15.9922693260499\\
0.5	16.3470844462236\\
0.6	16.4871640806582\\
0.7	16.5489387815016\\
0.8	16.5786332696136\\
0.9	16.5935286659942\\
};

\end{axis}

\begin{axis}[%
width=5.604in,
height=3.771in,
at={(0in,0in)},
scale only axis,
xmin=0,
xmax=1,
ymin=0,
ymax=1,
axis line style={draw=none},
ticks=none,
axis x line*=bottom,
axis y line*=left,
legend style={legend cell align=left, align=left, draw=white!15!black}
]
\end{axis}
		\end{tikzpicture}}
	\caption{Mean latency vs $p_A$ at $T=50$, $\theta=5$, $\lambda=0.5$, and $R=2$.}
\label{fig:delay}
\end{figure}
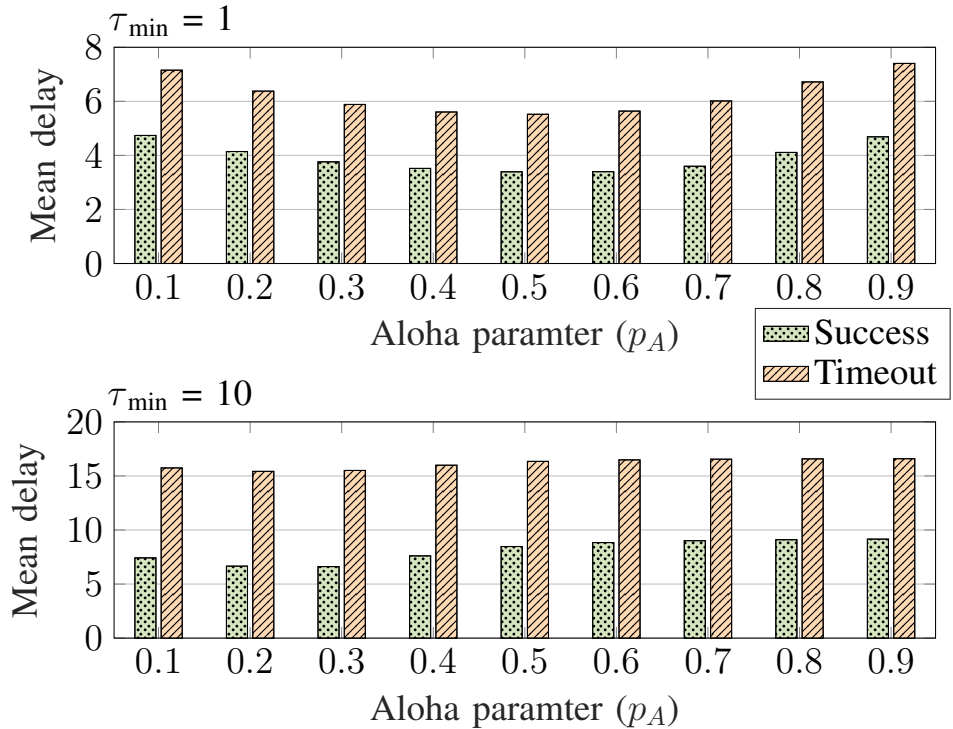

Fig.~\ref{fig:valid} shows the meta distribution of the TSP for different values of $p_A$ and $\tau_{\rm min}$. First, the close match between Monte Carlo simulations (i.e., marks) and analysis (i.e., curves) validates the developed model. The figure also shows the drastic impact of the Aloha parameter $p_A$ and minimum deadline $\tau_{\rm min}$ on the TSP. More strict $\tau_{\rm min}$ leads to higher probability of elapsed deadlines, which reliefs interference due to discarded packets and abandoned transmissions. Also, low values of $p_A$ implies conservative channel access for transmission attempts, which further reliefs interference and improves TSP.

Fig.~\ref{fig:absorb} shows the probability of absorption to success or timeout states for different values of Aloha parameter $p_A$ and minimum deadline $\tau_{\rm min}$. The figure reveals an optimal value of $p_A$ that depends on $\tau_{\rm min}$. Being too aggressive for channel access (i.e, high $p_A$) to catch the deadline leads to the counter effect of excessive transmissions failures due to deteriorating the TSP (cf. Fig~\ref{fig:valid}). On the other hand, being too conservative (i.e, low $p_A$) improves the TSP but at the expense of unnecessary backoffs that leads to missing the transmission deadlines. Hence, the value of $p_A$ should be carefully chosen based on the time criticality of packets deadlines. 

Surprisingly, Fig.~\ref{fig:absorb} shows the superior performance of strict packet deadlines in terms of the spatiotemporally averaged probability of successful transmission when compared to relaxed packet deadlines.  The effect of the aggregate network interference is the key to explain such counter-intuitive observation. Strict deadlines (i.e., lower $\tau_{min}$) relief the aggregate interference, and hence, improves the TSP. Therefore, the elapsed timeout of some devices help other devices to successfully deliver their packets. On the contrary, a relaxed timeout imposes aggressive channel access and leads to excessive transmission failures, which increases the probability of deadlines violations. 

Interestingly, strict deadlines also improve the mean latency for the successfully transmitted packets as shown in Fig.~\ref{fig:delay}. Exploiting the positive impact of strict deadline on the aggregate interference, a more aggressive channel access can be utilized. As shown in Figs.~\ref{fig:absorb} and \ref{fig:delay}, the optimal $p_{A}$ is higher for strict deadlines cases, which enhances both the probability of successful packet delivery as well as the latency of successfully delivered packets.

\savespaces
\section{Conclusion}\label{sec:conclude}
This paper develops a novel spatiotemporal model for an Aloha IoT network with asynchronous periodic traffic and hard transmission deadlines. A marked Poisson bipolar point process is used to model the IoT devices locations and the time offsets of their periodic packet generation. A discrete time absorbing Markov chain is utilized to model the temporal evolution of the packets until either successful transmission or deadline expiry. Stochastic geometry is utilized to account for the mutual interference among the active devices. To this end, the results reveal the drastic impact of the Aloha access probability and deadlines on the network performance. Interestingly, the results reveal the positive impact of strict deadline on both transmission success probability and latency.


\savespaces
\bibliographystyle{IEEEtran}
\bibliography{Aloha_deadline_Arxiv.bbl}

\begin{thebibliography}{10}
\providecommand{\url}[1]{#1}
\csname url@samestyle\endcsname
\providecommand{\newblock}{\relax}
\providecommand{\bibinfo}[2]{#2}
\providecommand{\BIBentrySTDinterwordspacing}{\spaceskip=0pt\relax}
\providecommand{\BIBentryALTinterwordstretchfactor}{4}
\providecommand{\BIBentryALTinterwordspacing}{\spaceskip=\fontdimen2\font plus
\BIBentryALTinterwordstretchfactor\fontdimen3\font minus
  \fontdimen4\font\relax}
\providecommand{\BIBforeignlanguage}[2]{{%
\expandafter\ifx\csname l@#1\endcsname\relax
\typeout{** WARNING: IEEEtran.bst: No hyphenation pattern has been}%
\typeout{** loaded for the language `#1'. Using the pattern for}%
\typeout{** the default language instead.}%
\else
\language=\csname l@#1\endcsname
\fi
#2}}
\providecommand{\BIBdecl}{\relax}
\BIBdecl

\bibitem{URLLC}
M.~{Bennis}, M.~{Debbah}, and H.~V. {Poor}, ``Ultrareliable and low-latency
  wireless communication: Tail, risk, and scale,'' \emph{Proc. IEEE}, vol. 106,
  no.~10, pp. 1834--1853, Oct 2018.

\bibitem{Hard_deadline}
R.~{Jurdi}, S.~R. {Khosravirad}, H.~{Viswanathan}, J.~G. {Andrews}, and R.~W.
  {Heath}, ``Outage of periodic downlink wireless networks with hard
  deadlines,'' \emph{IEEE Trans. on Commun.}, vol.~67, no.~2, pp. 1238--1253,
  Feb 2019.

\bibitem{First_Elsawy}
A.~{Bader}, H.~{ElSawy}, M.~{Gharbieh}, M.~{Alouini}, A.~{Adinoyi}, and
  F.~{Alshaalan}, ``First mile challenges for large-scale {IoT},'' vol.~55,
  no.~3, pp. 138--144, March 2017.

\bibitem{GhaElsBadAlo:17}
M.~{Gharbieh}, H.~{ElSawy}, A.~{Bader}, and M.~{Alouini}, ``Spatiotemporal
  stochastic modeling of {IoT} enabled cellular networks: Scalability and
  stability analysis,'' \emph{IEEE Trans. on Commun.}, vol.~65, no.~8, pp.
  3585--3600, Aug. 2017.

\bibitem{Chisci}
G.~{Chisci}, H.~{Elsawy}, A.~{Conti}, M.~{Alouini}, and M.~Z. {Win},
  ``Uncoordinated massive wireless networks: Spatiotemporal models and
  multiaccess strategies,'' \emph{IEEE/ACM Trans. on Netw.}, vol.~27, no.~3,
  pp. 918--931, June 2019.

\bibitem{Martin_queue}
Y.~{Zhong}, M.~{Haenggi}, T.~Q.~S. {Quek}, and W.~{Zhang}, ``On the stability
  of static {Poisson} networks under random access,'' vol.~64, no.~7, pp.
  2985--2998, July 2016.

\bibitem{Gharbieh}
M.~{Gharbieh}, H.~{ElSawy}, H.~{Yang}, A.~{Bader}, and M.~{Alouini},
  ``Spatiotemporal model for uplink {IoT} traffic: Scheduling and random access
  paradox,'' \emph{IEEE Trans. on Wireless Commun.}, vol.~17, no.~12, pp.
  8357--8372, Dec 2018.

\bibitem{Tony}
Y.~{Zhong}, T.~Q.~S. {Quek}, and X.~{Ge}, ``Heterogeneous cellular networks
  with spatio-temporal traffic: Delay analysis and scheduling,'' vol.~35,
  no.~6, pp. 1373--1386, Jun. 2017.

\bibitem{Deng}
N.~{Jiang}, Y.~{Deng}, X.~{Kang}, and A.~{Nallanathan}, ``Random access
  analysis for massive {IoT} networks under a new spatio-temporal model: A
  stochastic geometry approach,'' \emph{IEEE Trans. on Commun.}, vol.~66,
  no.~11, pp. 5788--5803, Nov 2018.

\bibitem{Nardelli}
P.~S. {Dester}, P.~{Cardieri}, P.~H.~J. {Nardelli}, and J.~M.~C. {Brito},
  ``Performance analysis and optimization of a $n$ -class bipolar network,''
  \emph{IEEE Access}, vol.~7, pp. 135\,118--135\,132, 2019.

\bibitem{AoI}
M.~{Emara}, H.~{ElSawy}, and G.~{Bauch}, ``A spatiotemporal model for peak
  {AoI} in uplink {IoT} networks: Time vs event-triggered traffic,'' \emph{IEEE
  Internet of Things Journal}, pp. 1--15, 2020.

\bibitem{Sheng}
L.~{Liu}, M.~{Sheng}, J.~{Liu}, Y.~{Dai}, and J.~{Li}, ``Stable throughput
  region and average delay analysis of uplink {NOMA} systems with unsaturated
  traffic,'' \emph{IEEE Trans. on Commun.}, vol.~67, no.~12, pp. 8475--8488,
  Dec 2019.

\bibitem{Fatma}
F.~{Benkhelifa}, H.~{ElSawy}, J.~A. {Mccann}, and M.~{Alouini}, ``Recycling
  cellular energy for self-sustainable {IoT} networks: {A} spatiotemporal
  study,'' \emph{IEEE Trans. Wireless Commun.}, vol.~19, no.~4, pp. 2699--2712,
  2020.

\bibitem{Hae:J16}
M.~Haenggi, ``The meta distribution of the {SIR} in {Poisson} bipolar and
  cellular networks,'' vol.~15, no.~4, pp. 2577--2589, Apr. 2016.

\bibitem{alfa}
S.~Chakravarthy and A.~S. Alfa, \emph{Matrix-analytic methods in stochastic
  models}.\hskip 1em plus 0.5em minus 0.4em\relax CRC Press, 1996.

\end{thebibliography}

\end{document}